%% file: PRD_main.tex
\documentclass[aps, prd, twocolumn, nofootinbib, floatfix, superscriptaddress]{revtex4-2}
\usepackage{subfigure}
\usepackage{graphicx}
\usepackage{dcolumn}
\usepackage{epsfig}
\usepackage{amsmath}
\usepackage{amsfonts}
\usepackage{graphicx}
\usepackage{amssymb}
\usepackage{amssymb,amsmath,amsfonts}
\usepackage{xfrac}
\usepackage{color}
\usepackage{tikz}
\usepackage[utf8]{inputenc}

\usepackage{tcolorbox}
\usepackage{hyperref}
\hypersetup{colorlinks, citecolor=bluscuro, linkcolor=black, urlcolor=bluscuro}
\definecolor{rossos}{cmyk}{0,1,1,0.55}
\definecolor{bluscuro}{rgb}{0.15, 0.2, .85}
\definecolor{bluchiaro}{cmyk}{1,.3,0.,0.1}

\input{newcommands}

\newcommand{\be}{\begin{equation}}

\begin{document}

\title{Scalar induced gravitational waves in modified teleparallel gravity theories}

\author{Charalampos Tzerefos}
\email{chtzeref@phys.uoa.gr}
\affiliation{ \mbox{National Observatory of Athens, Lofos Nymfon, 11852 Athens, Greece} }

\affiliation{\mbox{Department of Physics, National \& Kapodistrian University of Athens, Zografou Campus GR 157 73, Athens, Greece}}

\author{Theodoros Papanikolaou}
\email{theodoros.papanikolaou@noa.gr}
\affiliation{ \mbox{National Observatory of Athens, Lofos Nymfon, 11852 Athens, Greece} }

\author{Emmanuel N. Saridakis}
\email{msaridak@noa.gr}
\affiliation{ \mbox{National Observatory of Athens, Lofos Nymfon, 11852 Athens, Greece} }
\affiliation{CAS Key Laboratory for Researches in Galaxies and Cosmology, 
Department of Astronomy, \\
University of Science and Technology of China, Hefei, 
Anhui 230026, P.R. China}
\affiliation{\mbox{Departamento de Matem\'{a}ticas, Universidad Cat\'{o}lica del 
Norte, 
Avda.
Angamos 0610, Casilla 1280 Antofagasta, Chile}}

\author{Spyros Basilakos}
\email{svasil@Academyofathens.gr}
\affiliation{ \mbox{National Observatory of Athens, Lofos Nymfon, 11852 Athens, Greece} }
\affiliation{ \mbox{Academy of Athens, Research Center for Astronomy and Applied Mathematics, Soranou Efesiou 4, 11527, Athens, Greece} }
\affiliation{ \mbox{School of Sciences, European University Cyprus, Diogenes Street, Engomi, 1516 Nicosia, Cyprus} }

\begin{abstract} 
Primordial black  holes (PBHs) forming out of the collapse of enhanced 
cosmological perturbations provide   access to the early Universe through their 
associated observational signatures. In particular, enhanced cosmological 
perturbations collapsing to form PBHs are responsible for the generation of a 
stochastic gravitational-wave background (SGWB) induced by second-order 
gravitational interactions, usually called   scalar induced gravitational waves 
(SIGWs). This SGWB is sensitive to the underlying gravitational theory; hence it 
can be used as a novel tool to test the standard paradigm of gravity and 
constrain possible deviations from general relativity. In this work, we study 
the aforementioned GW signal within   modified teleparallel gravity theories, 
developing a formalism for the derivation of the GW spectral abundance within 
any form of   gravitational action. At the end, working within viable  $f(T, 
\phi)$ models without matter-gravity couplings, and accounting for the effect 
of mono-parametric $f(T)$ gravity at the level of the source and the propagation 
of the tensor perturbations, we show that the respective GW signal is 
indistinguishable from that within GR. Interestingly, we find that in order to 
break the degeneracy between different $f(T)$ theories through the portal of 
SIGWs one should necessarily consider non-minimal matter-gravity couplings at 
the level of the gravitational action.
\noindent

\end{abstract}

\maketitle
\section{Introduction}

Primordial black holes (PBHs), firstly introduced in the 
early '70s~\cite{1967SvA....10..602Z, Carr:1974nx,1975ApJ...201....1C},  have 
gained lot of attention within the scientific community since they can naturally 
address a number of fundamental issues of modern cosmology. In particular, they 
can potentially account for a part or the totality of dark 
matter~\cite{Chapline:1975ojl,Clesse:2017bsw} and explain the large-scale 
structure formation through   Poisson 
fluctuations~\cite{Meszaros:1975ef,Afshordi:2003zb}. At the same time, depending 
on their mass they can give rise to a very rich phenomenology from 
the early universe up to late times~\cite{Carr:2020gox}.

Meanwhile, PBHs are connected with numerous gravitational-wave (GW) 
signals~\cite{Sasaki:2018dmp,Zhou:2020kkf}. Since the detection of the first GW 
signal in 2015, there has been a lot of progress in the literature connecting 
PBHs with the GWs. More specifically, there have been extensively studied 
GWs from PBH merging events~\cite{Nakamura:1997sm, Ioka:1998nz, Zaballa:2006kh,
Raidal:2017mfl,Cotner:2019ykd}, GWs which are induced from enhanced scalar 
perturbations collapsing to PBHs due to second-order gravitational 
interactions~\cite{Bugaev:2009zh, Saito_2009, Nakama_2015} 
[See~\cite{Domenech:2021ztg} for a recent review] as well as GWs induced by 
Poisson PBH energy density perturbations 
themselves~\cite{Papanikolaou:2020qtd,Domenech:2020ssp,Papanikolaou:2022chm}.

In particular, the portal of scalar induced gravitational waves (SIGWs) 
constitutes an active field of research since they can give us access to the 
thermal history of the 
Universe~\cite{Li:2017drr,Cai:2019jah,Domenech:2020kqm,Cai:2020ovp} and in 
particular on the conditions that prevailed in the early Universe, namely during 
cosmic inflation~\cite{Bugaev:2009zh,Saito_2009,Fumagalli:2020nvq,Fumagalli:2021mpc,Cai:2019bmk,Cai:2019amo} and  
reheating~\cite{Domenech:2019quo} during which all the known particles are 
considered to have been produced. Interestingly enough, through the portal of 
SIGWs one can have access to very small scales which are poorly constrained and 
are otherwise inaccessible with Cosmic Microwave Background (CMB) and Large 
Scale Structure (LSS) probes~\cite{Domenech:2021ztg} while very encouragingly, 
the typical frequency of such primordial GWs lie well within the frequency 
detection band of future GW detectors such as the Einstein Telescope 
(ET)~\cite{Maggiore:2019uih}, the Laser Inferometer Space Antenna 
(LISA)~\cite{Caprini:2015zlo,Karnesis:2022vdp} and the Square Kilometer Arrays 
(SKA)~\cite{Janssen:2014dka}.

Up to now, the majority of the works in the literature investigated the  
aforementioned GW signal within the context of general relativity (GR). However, 
there are many theoretical as well as phenomenological reasons which point 
toward a different gravity paradigm in order to account indicatively for the 
renormalizability issues of classical 
gravity~\cite{Stelle:1976gc,Addazi:2021xuf} and explain the two phases of the 
Universe's 
accelerated expansion, namely the early-time, inflationary one 
\cite{Nojiri:2010wj,Martin:2013tda}, and/or the late-time, dark-energy one 
\cite{CANTATA:2021ktz,Capozziello:2011et,Cai:2015emx}.  In view of these 
arguments, SIGWs are promoted as a novel portal to test and constrain the 
underlying gravity theory. 

Recently, there has been an increased scientific activity toward  this direction 
through the study of primordial SIGWs within curvature formulations of 
gravity~\cite{Lin:2020goi,Chen:2021nio,Kawai:2021edk,Lin:2021vwc,Papanikolaou:2021uhe,Ivanov:2021chn,Zhang:2021rqs,Yi:2022anu,Cheong:2022gfc,Feng:2023veu,Zhang:2022xmm,Arya:2023pod}. In the present work, we study for the first time to the 
best of our knowledge the primordial SIGW portal within the context of a torsional 
formulation of gravity where the gravitational Lagrangian is promoted to an 
integral of a function of the torsion scalar $T$ containing potentially 
couplings between the gravity and the matter sectors of the 
Universe~\cite{Aldrovandi:2013wha, 
Krssak:2018ywd,Cai:2015emx,Bengochea:2008gz,Linder:2010py,
Chen:2010va,
 Zheng:2010am,Bamba:2010wb,Cai:2011tc,
Capozziello:2011hj,  
 Bamba:2013jqa,
Li:2013xea,Ong:2013qja,  
Bamba:2016gbu,Malekjani:2016mtm,Farrugia:2016qqe,
Bahamonde:2017wwk,Karpathopoulos:2017arc,Abedi:2018lkr,DAgostino:2018ngy,
Krssak:2018ywd,
Iosifidis:2018zwo,
Chakrabarti:2019bed, DavoodSadatian:2019pvq,Yan:2019gbw,Wang:2020zfv, 
Bose:2020xdz, Ren:2021tfi,Escamilla-Rivera:2021xql}. In particular, by  studying 
the effect of modified teleparallel gravity theories at the level of the source 
and the propagation of the SIGWs we examine under which conditions one can 
detect a distinctive deviation from the case of classical gravity.

The paper is structured as follows: In \Sec{teleparallel} we review the  
fundamentals of the torsional formulation of gravity studying its background and 
perturbation behavior and specifying as well viable $f(T,\phi)$ gravity models within 
which we study the SIGW signal. Then, in \Sec{SIGW} we present the basics of the 
SIGWs by studying at the same time the effect of modified teleparallel gravity 
(MTG) theories at the level of the source and the propagation of the GWs. 
Furthermore, we deduce the necessary conditions so as to see a distinctive SIGW 
signature within MTG theories compared to classical gravity. Finally, 
\Sec{sec:Conclusions} is devoted to conclusions.

\section{General framework of modified teleparallel 
gravity theories}
\label{teleparallel}
\subsection{Teleparallel gravity} 

Teleparallel Gravity (TG) is an alternative formulation of gravity based on 
torsion \cite{DeAndrade:2000sf, Arcos:2004tzt, Pereira:2019woq}. The dynamical 
variable of TG is the tetrad field, $\mathbf{e}^{}_{A}(x^\mu)$ and it connects 
the spacetime metric $g_{\mu\nu}$ and the Minkowski tangent space metric $\eta 
_{AB}^{}=\text{diag}\,(-1,1,1,1)$ through the following relation:
\begin{equation}  \label{metrics}     
 g_{\mu\nu}=e^{A}_{~\mu}\,e^{B}_{~\nu}\,\eta_{AB}^{}\,,
 \end{equation} where  Greek and Latin indices run in coordinate 
and tangent space respectively and $e^{A}_{~\mu}$ are the tetrad components  
which satisfy the orthonormality conditions  $e^{A}_{~\mu} 
e_{A}^{~\nu}=\delta^{\nu}_{\mu}$ and $e^{A}_{~\mu} e_{B}^{~\mu}=\delta^{A}_{B}$, 
with $e_{B}^{~\mu}$ being the inverse components.

 Due to relation \eqref{metrics}, the tetrad fields are only determined up to 
transformations of the six-parameter Lorentz group. To ensure the covariance of 
the theory one needs to introduce a Lorentz or spin connection 
\cite{Weinberg:1972kfs}, which can be written as
 \beq
\omega^{A}_{~B \mu}=\Lambda^{A}_{~D}(x) \partial_{\mu}{\Lambda_{B}^{~D}(x)},
\label{spin_TG}
\eeq
 with $\Lambda^{A}_{~D}(x)$ being a local (point-dependent)  Lorentz 
transformation~\cite{Aldrovandi:1996ke}. 
 TG is characterised by the choice to formulate gravity in a particular class of 
frames (called proper frames) for which the spin connection is flat, i.e. 
$\omega^{A}_{~B \mu}=0$. This choice is facilitated by the local Lorentz  
invariance of TG. The corresponding spacetime-indexed connection which is the 
so-called Weitzenb\"{o}ck connection \cite{Aldrovandi:2013wha} is the following:
\beq
\Gamma^{\rho}_{~\mu\nu}=e_{A}^{~\rho} 
\left(\partial_{\mu}{e^{A}_{~\nu}}+\omega^{A}_{~B \mu}e^{B}_{~\nu}\right) 
\Rightarrow \overset{\mathbf{w}}{\Gamma}^\lambda_{\nu\mu}\equiv e^\lambda_A\:
\partial_\mu
e^A_\nu.
\eeq
The action functional of TG is defined by
\beq
S=-\frac{\Mp^2}{2}\int{\mathrm{d}^{4}x\,e\,{T}},
\label{action_TG}
\eeq with $e=\det{\left(e^{A}_{~\mu}\right)}=\sqrt{-g}$ and  $\Mp^2\equiv 
\left(8\pi G\right)^{-1}$ being the reduced Planck mass. The torsion scalar $T$ 
is defined by 
\beq
T= S_{\rho}^{~\mu\nu}\,T^{\rho}_{~\mu\nu},
\label{ScalarT}
\eeq
with $ T^{\rho}_{~\mu\nu}$ being the components of the torsion tensor defined by 
\begin{equation} \label{Def_Torsion}   
 T^{\rho}_{~\mu\nu}\equiv e_{A}^{~\rho}\left[\partial_{\mu}e^{A}_{~\nu}
 -\partial_{\nu}e^{A}_{~\mu}+\omega^{A}_{~B\mu}\,e^{B}_{~\nu},
 -\omega^{A}_{~B\nu}\,e^{B}_{~\mu}\right]\,
\end{equation}
and $S_{\rho}^{~\mu\nu}$ being the so-called super-potential which reads as
\begin{equation} \label{Superpotential} 
S_{\rho}^{~\mu\nu}\equiv\frac{1}{2}\left(K^{\mu\nu}_{~~\rho}+\delta^{\mu}_{~\rho
}  
\,T^{\theta\nu}_{~~\theta}-\delta^{\nu}_{~\rho}\,T^{\theta\mu}_{~~\theta}
\right)\,,
\end{equation} with $K^{\mu\nu}_{~~\rho}$ standing for the contortion tensor 
defined by
\begin{equation}  \label{Contortion}
 K^{\mu\nu}_{~~\rho}\equiv -\frac{1}{2}\left(T^{\mu\nu}_{~~\rho}
 -T^{\nu\mu}_{~~\rho}-T_{\rho}^{~\mu\nu}\right).
\end{equation} 
 The Weitzenb\"{o}ck connection of TG and the Levi-Civita connection of GR, 
$\bar{\Gamma}^{\rho}_{~\mu\nu}$,  are related as follows
\beq
\Gamma^{\rho}_{~\mu\nu}=\bar{\Gamma}^{\rho}_{~\mu\nu}+K^{\rho}_{~\mu\nu}.
\label{Gamma_relation}
\eeq
Consequently, it can be shown that 
\beq
T=-R-2e^{-1} \partial_{\mu}{\left(e T^{\nu \mu}_{~~~\nu}~\right)},
\eeq
with $R$ being the curvature scalar of the Levi-Civita connection 
\cite{Bahamonde:2021gfp}.  Therefore, TG and GR are equivalent theories at the 
level of the field equations.

However, when one extends TG by introducing a non-minimally coupled matter 
field, for instance a scalar field  
\cite{Geng:2011aj,Otalora:2013tba,Otalora:2013dsa,Otalora:2014aoa,
Skugoreva:2014ena}, or by adding into the action non-linear terms in the torsion 
scalar $T$, as for example in $f(T)$ gravity 
\cite{Bengochea:2008gz,Linder:2010py,Li:2011wu,Gonzalez-Espinoza:2018gyl}, one 
obtains new classes of modified gravity theories with interesting phenomenology 
which are not equivalent to their corresponding curvature based counterparts 
\cite{Cai:2015emx}.

In the following, we shall briefly present the generation of primordial density 
fluctuations  in the framework of generalized teleparallel scalar-torsion 
gravity theories following \cite{GONZALEZESPINOZA2020135696}. 

\subsection{Generalized scalar-torsion gravity}
\subsubsection{Field equations}

 By extending the gravitational sector to an arbitrary function of $T$ and 
$\phi$, the corresponding action functional of the generalized scalar-torsion 
gravity is given by 
\cite{Geng:2011aj,Otalora:2013tba,Otalora:2013dsa,Otalora:2014aoa,
Skugoreva:2014ena}
\begin{equation} 
 S=\int \mathrm{d}^{4}x\,e\,\left[ f(T,\phi)+ P(\phi)X \right] \label{gaction},
\end{equation}
with $X$ being the so-called canonical kinetic term defined by  $X\equiv- 
\partial_{\mu}{\phi}\partial^{\mu}{\phi}/2$. Teleparallel gravity with a scalar 
field potential $V(\phi) $ is recovered when $f(T,\phi)=-\Mp^2 T/2-V(\phi)$. 

The corresponding field equations are obtained by varying this action with 
respect to  the tetrad field $e^{A}_{~\mu}$ \cite{Gonzalez_Espinoza_2021}:
\begin{align}
\begin{split}
f_{,T} G_{\mu \nu} & +S_{\mu \nu}{}^{\rho} \partial_{\rho} f_{,T}+\frac{1}{4} g_{\mu \nu}\left(f-T f_{,T}\right) \\ & +
\frac{P}{4}\left(g_{\mu \nu} X+\partial_{\mu}\phi \partial_{\nu}\phi\right)=0,
\end{split}
\label{FieldEquations}
\end{align} where a comma denotes partial differentiation, here with respect to 
$T$.  These equations have been expressed in a general coordinate basis with 
$G^{\mu}_{~\nu}=e_{A}^{~\mu} G^{A}_{~\nu}$ being the Einstein tensor and  
$G_{A}^{~\mu}\equiv e^{-1}\partial_{\nu}\left(e e_{A}^{~\sigma} 
S_{\sigma}^{~\mu\nu}\right)-e_{A}^{~\sigma} T^{\lambda}_{~\rho 
\sigma}S_{\lambda}^{~\rho \mu}+e_{B}^{~\lambda} S_{\lambda}^{~\rho 
\mu}\omega^{B}_{~A \rho}+\frac{1}{4}e_{A}^{~\mu} T$. 

 It is important to point out that the action (\ref{gaction}) is not locally 
Lorentz  invariant \cite{Sotiriou:2010mv,Li:2010cg}. One can easily see this by 
performing an infinitesimal Lorentz transformation to the tetrads as follows: 
$e'^{A}_{~\mu}=e^{A}_{~\mu}+\xi_{B}^{~A}e^{B}_{~\mu}$, with $\xi^{A B}=-\xi^{B 
A}$. The effect of this transformation on the action \label{action1} is  
\begin{equation}
\delta{S}=\int d^{4}{x}e \partial_{\rho} f_{,T} S_{\mu \nu}^{\rho} e_{A}^{~\mu}  
e_{B}^{~\nu}\xi^{A B}. \, \, 
\end{equation} 
Now if one demands that this action is invariant, that is $\delta{S}=0$ for  
arbitrary $\xi^{A B}$, the following equation needs to be satisfied
\begin{equation}
\partial_{\rho} f_{,T} S_{\left[\mu \nu\right]}{}^{\rho}=0.
\label{Constr}
\end{equation} 
Thus, since equation (\ref{Constr}) is not satisfied in general, the action  
(\ref{gaction}) is not Lorentz invariant locally. For the special case of TG, 
$f\sim T \Rightarrow \partial_{\rho} f_{,T}=0$, therefore (\ref{Constr}) is 
satisfied. 

\subsubsection{Cosmological framework}

To apply this general formulation into a cosmological setting, one needs to  
impose the standard flat, homogeneous and isotropic  Friedmann - Lema\^itre - 
Robertson -Walker (FLRW) geometry
\begin{equation}
\mathrm{d}s^2=-\mathrm{d}t^2+a(t)^2\,\delta_{ij} \mathrm{d}x^i \mathrm{d}x^j \,,
\label{FRWMetric}
\end{equation}
which corresponds to the following tetrads
\begin{equation}
\label{veirbFRW}
e^A_{~\mu}={\rm
diag}(1,a(t),a(t),a(t)),
\end{equation}
with $a(t)$ being the scale factor. By substituting the tetrad field 
\eqref{veirbFRW} \footnote{It is worth noting that due to the violation of 
local Lorentz invariance in general MTG theories 
\cite{Sotiriou:2010mv}, there is the following complication: the 
gravitational field equations and their tetrad solutions become dependent on the 
corresponding spin connection. Consequently, one needs a way to retrieve the 
corresponding spin connection associated with each tetrad field in order to 
properly solve the field equations. For our FLRW setting, 
 it has been shown that our chosen tetrad \eqref{veirbFRW} is a proper tetrad,  
which implies that its corresponding spin connection is the vanishing spin 
connection leading to physically meaningful results \cite{Krssak:2015oua}.} into 
the field equations \eqref{FieldEquations} one obtains the following background 
equations
\begin{eqnarray}
    \label{Fr1}
 f(T,\phi) - P(\phi) X - 2 T f_{,T}=0, \\
\label{Fr2}
 f(T,\phi) + P(\phi) X - 2 T f_{,T} - 4 \dot{H} f_{,T} - 4 H \dot{f}_{,T}=0, \\
\label{phi}
 - P_{,\phi} X - 3 P(\phi) H \dot{\phi} - P(\phi) \ddot{\phi} + f_{,\phi}=0,
\end{eqnarray}
where $H\equiv \dot{a}/a$ is the Hubble parameter and a dot denotes derivative 
with respect to $t$. Additionally, from equation \eqref{ScalarT} one obtains 
$T=6 H^2$. 

In order to describe slow-roll inflation, one needs to introduce the following 
 slow-roll parameters
\begin{eqnarray}
&& \epsilon\equiv -\dfrac{\dot{H}}{H^2},\:\: \delta_{P X} \equiv - \dfrac{P(\phi) X}{2 H^{2} f_{,T}},\:\:\ \delta_{f_{,T}} \equiv \dfrac{\dot{f}_{,T}}{f_{,T} H},
 \label{slowpara1}
\end{eqnarray}
such as that from equations \eqref{Fr1} and \eqref{Fr2} one can write $\epsilon$ as
\begin{equation}
\epsilon =  \delta_{PX} + \delta_{f_{,T}}.
\label{slow_roll_eps}
\end{equation}
Furthermore, it is useful to split the parameter $\delta_{f_{,T}}$ as
\beq
\delta_{f_{,T}}=\delta_{f\dot{H}}+\delta_{fX},
\label{deltafT}
\eeq by defining
\beq
\delta_{f\dot{H}}\equiv \frac{f_{,TT} \dot{T}}{H f_{,T}},\:\:\:\: \delta_{fX}\equiv \frac{f_{,T\phi}\dot{\phi}}{H f_{,T}}.
\eeq
Therefore, from expressions \eqref{slowpara1} and \eqref{slow_roll_eps}, one 
can obtain the following relations
\begin{eqnarray}
 &&\delta_{f\dot{H}}=-\frac{2 \mu}{1+2\mu}\left(\delta_{PX}+\delta_{fX}\right), \\ 
&&\delta_{f_{,T}}=\frac{1}{1+2\mu} \left(\delta_{fX}-2 \mu \delta_{PX}\right), \\ 
&&\epsilon=\frac{1}{1+2\mu}\left(\delta_{PX}+\delta_{fX}\right)  \label{e},
\end{eqnarray} where we have defined $\mu\equiv T f_{,TT}/f_{,T}$ in  analogy 
with the deviation parameter of the (curvature based) modified gravity theories 
\cite{DeFelice:2010aj}.

\subsection{Scalar perturbations}

In order to describe scalar perturbations, it is convenient to employ  the 
Arnowitt-Deser-Misner (ADM) decomposition of the tetrad field \cite{Wu:2011kh} 
where
\begin{eqnarray}
\begin{split}
& e^{0}_{~\mu} = \left(N,\textbf{0}\right),\:\:\:\: e^{a}_{~\mu}=\left(N^{a},h^{a}_{~i}\right)\label{ADM1},\:\:\:\: 
\\ &
 e_{0}^{~\mu}=\left(1/N,-N^{i}/N\right),\:\:\:\: e_{a}^{~\mu}=\left(0, h_{a}^{~i}\right)\label{ADM2},
 \end{split}
\end{eqnarray} 
with $N$ being the lapse function and $N^i$ the shift vector, which is defined by $N^{i}\equiv h_{a}^{~i}  N^{a}$ and $h^{a}_{~i}$ being the induced tetrad field satifying the orthonormality condition, i.e. $h^{a}_{~j} h_{a}^{~i}=\delta^{i}_{j}$.

Choosing to work within the uniform field gauge, or otherwise called  comoving 
gauge, i.e. $\delta \phi=0$, a convenient ansatz for the lapse function, the 
shift vector and the induced tetrad fields is 
\beq
N=1+A,\:\:\:\: N^{a}=a^{-1} e^{-\mathcal{R}}\delta^{a}_{~i}  
\partial^{i}{\psi},\:\:\:\: h^{a}_{~i}=a 
e^{\mathcal{R}}\delta^{a}_{~j}\delta^{j}_{~i},
\label{Uniform_Field_Gauge}
\eeq which gives rise to the corresponding perturbed metric  
\cite{DeFelice:2011uc}
\beq
\begin{split}
 \mathrm{d}s^2  = & -\left[\left(1+A \right)^2-a^{-2}e^{-2\mathcal{R}} 
\left(\partial \psi\right)^2\right]\mathrm{d}t^2  
\\ & +2\partial_{i}{\psi}\mathrm{d}t \mathrm{d}x^{i} + a^2 e^{2\mathcal{R}}\delta_{i 
j} \mathrm{d}x^{i} \mathrm{d}x^{j}.
\end{split}
\eeq

Now one needs to expand the action (\ref{gaction}) up to  second order in the 
perturbation variables of the perturbed tetrad (\ref{Uniform_Field_Gauge}). In 
order to accomplish this, one needs to address the fact that the action is not 
Lorentz invariant locally. The standard procedure for that essentially consists 
in adding the additional six Lorentz degrees of freedom, which arise because of 
the Lorentz violation, directly into the perturbed tetrad field 
\eqref{Uniform_Field_Gauge} \cite{Izumi:2012qj,Golovnev:2018wbh}. Afterwards, 
once a particular perturbed tetrad frame is chosen, these extra modes can be 
absorbed  into Goldstone modes of the Lorentz symmetry breaking, by performing a 
Lorentz rotation of the tetrad field \cite{Bluhm_2005, Bluhm_2008}. After this procedure, a new massive term is generated and the corresponding action is

\begin{equation}
S^{(2)} =\frac{1}{2}\int \mathrm{d}\tau \mathrm{d}^{3}x \left[ (v')^{2} -  (\partial v)^{2} - M^{2}  v^{2} \right],
\label{SOA_v}
\end{equation} 
where we defined the usual Mukhanov-Sasaki (MS) variable 
\beq
v \equiv z \mathcal{R}, \, \text{with} \, \,  z^{2} \equiv 2 a^{2} Q_{s}\,\,\, \text{and} \,\, \, Q_s \equiv \frac{PX}{H^2}, \label{eq:z_and_Qs}
\eeq
where the prime denotes differentiation with respect to the conformal time  $\tau$ defined by $d\tau\equiv dt/a$. The $M$   is an effective mass parameter 
defined by 
\beq
M^{2} \equiv a^{2} m^{2} - \dfrac{z''}{z}, 
\eeq 
where $m^2=3 H^2 \eta_{\mathcal{R}}$ and $\eta_{\mathcal{R}}$ is given by 
\begin{eqnarray}
\eta_{\mathcal{R}}=\frac{m^2}{3 H^2}=\delta_{f_{,T}}\left[1 + \left(1+\frac{\delta_{fX}}{\delta_{PX}}\right)\dfrac{\delta_{f_{,T}}}{\delta_{f\dot{H}}} \right].
\label{mass_term} 
\end{eqnarray}
The parameter $m$ is a new explicit mass term, which arises due to the effects 
of local Lorentz-symmetry breaking mentioned earlier. 

By varying  the action \eqref{SOA_v} and using the Fourier expansion of the MS 
variable 
\begin{equation}
v( \tau, \textbf{x} ) = \int \dfrac{\text{d}^{3} k}{(2 \pi)^{3}}  v_{\textbf{k}} 
(\tau) e^{i \textbf{k}. \textbf{x}},
\end{equation} 
one obtains the following field equation 
\begin{equation}
v''_{k} + \left( k^{2} + M^{2}\right) v_{k} = 0,
\label{MukhSassa_Eq}
\end{equation} 
which is the corresponding Mukhanov-Sasaki equation within the modified 
teleparallel gravity setup. Given now that the MS variable $v$ is related to the 
comoving curvature perturbation $\mathcal{R}$ as $v=z \mathcal{R}$ where $z$ is 
given by \Eq{eq:z_and_Qs}, one can rewrite \Eq{MukhSassa_Eq} in terms of the 
comoving curvature perturbation $\mathcal{R}$ as follows: \footnote{In our 
numerical implementation, we used the e-fold number $N$ defined as $N\equiv \ln 
a$ as our time variable.}

\beq
\mathcal{R}^{\prime\prime}_k +  2\frac{z^\prime}{z} \mathcal{R}^\prime_k +  \left( k^2 + 
a^2m^2\right)\mathcal{R}_k = 0 \label{MK_conformal}.
\eeq

\subsection{Tensor perturbations}

To describe the tensor perturbations we  shall adopt again the uniform field 
gauge, $\delta \phi=0$, so from our earlier ADM decomposition of the tetrad 
field from equations \eqref{ADM1} we get that \cite{Wu:2011kh, 
GONZALEZESPINOZA2020135696} 
\begin{align}
N=1,\:\:\: N^{a}=0, \:\:\: h^{a}_{~i}=a(\delta^{a}_{~i}+\frac{1}{2}\gamma^{a}_{~i}) \label{tensor_tetrads}.
\end{align}
Then we can define the induced $3-$metric 
\begin{eqnarray}
g_{ij}=\eta_{ab} h^{a}_{~i} h^{b}_{~j}=a^2 \left[\delta_{i j}+h_{i j}+\frac{1}{4} \gamma_{k i} \gamma^{k}_{~j}\right],
\end{eqnarray} where we  defined the spatial tensor modes by 
\beq
h_{i j}=\frac{1}{2}\eta_{ab}\left(\delta^{a}_{~i}\gamma^{b}_{~j}+\delta^{b}_{~j}\gamma^{a}_{~j}\right)=\frac{1}{2}\left(\gamma_{i j}+\gamma_{j i}\right), 
\eeq
with $\gamma^{a}_{~j}=\gamma^{i}_{~j}\delta^{a}_{~i}$. It is illustrating to 
decompose the tensor $\gamma_{i j}$ into its symmetric and anti-symmetric part 
$\gamma_{i j}=\gamma_{\left(i,j\right)}+\gamma_{\left[i,j\right]}$. The 
symmetric part $h_{ij}=\gamma_{\left(i,j\right)}$ is gauge invariant 
\cite{baumann_2022} and satisfies the transverse and traceless conditions, i.e. 
$\partial^{i}h_{ij}=h^{i}_{i}=0$, while the antisymmetric part matches the gauge 
degrees of freedom in the local Lorentz invariant theory. We now need to 
substitute the tetrad fields (\ref{tensor_tetrads}) into the action 
(\ref{gaction}) and expand to second order in the tensor modes. For this 
purpose, one can neglect the $\gamma^2$ term  since it contributes only in cubic 
calculations of the Lagrangian \cite{Maldacena:2002vr}.

Consequently, the respective  second-order gravitational action for the tensor 
perturbations can be recast as:
\begin{equation}
S^{(2)}_\mathrm{T} = \int \mathrm{d}\tau \mathrm{d}^{3}x a^2 Q_T \left[ (h'_\lambda)^{2} -  (\partial h_\lambda)^{2}  \right],
\label{eq:S_2_T}
\end{equation}
with $Q_T$ being defined by $Q_T\equiv -f_{,T}/2$ and $\lambda = (+)$ or $(\times) $ 
accounting for two polarisation states of the tensor modes. At the end, 
minimising the aforementioned second-order action for the tensor modes  and 
Fourier transforming $h_\lambda$ one obtains the following equation of motion 
for $h_\boldmathsymbol{k}^{\lambda}$
\beq
\label{Tensor Eq. of Motion_no_source}
h_\boldmathsymbol{k}^{\lambda,\prime\prime} + 
2\mathcal{H} (1-\gamma_T)  h_\boldmathsymbol{k}^{\lambda,\prime} + k^{2}  
h^\lambda_\boldmathsymbol{k} = 0,
\eeq
with 
\beq
\label{eq:gamma_T}
 \gamma_T \equiv -\frac{f_T'}{2\mathcal{H} f_T}.
\eeq
\subsection{Specific $f(T,\phi)$ gravity models} \label{sec:f(T) models}

For concreteness, we will work with specific $f(T,\phi)$ gravity models with 
canonical kinetic terms, namely with $P(\phi) = 1$,  and without explicit 
non-minimal matter-gravity couplings, i.e. with $f_{,T\phi}=0$. Therefore, we shall work with models of the form $f(T,\phi) = f(T) + X - V(\phi)$. In the 
following we provide  the mono-parametric $f(T)$ 
gravity models that we will use.

\subsubsection{Power-law model}

The power-law model \cite{Bengochea:2008gz}
(hereafter $f_{1}$ model), in which 
\begin{equation}
f(T)=-\frac{\Mp^2}{2}\left(T+\alpha T^{\beta}\right) ,
\label{powermod}
\end{equation} 
with
\begin{eqnarray}
\alpha=(6H_0^2)^{1-\beta}\frac{\Omega_{F0}}{2\beta-1},
\end{eqnarray}
where $\Omega_{F0}=1-\Omega_{m0}-\Omega_{r0}$. According to  
observational constraints for $\beta$ one has that 
$-0.3<\beta<0.3$ \cite{Nesseris:2013jea,Nunes:2018evm,Anagnostopoulos:2019miu}
and the GR case is recovered for 
$\beta\rightarrow0$.

\subsubsection{Exponential model}
The  exponential model (hereafter $f_{2}$) \cite{Nesseris:2013jea}:
\begin{eqnarray}
f(T)=-\Mp^2/2 \left[T+\alpha T_{0}\left(1-e^{-T/(\beta T_{0})}\right)\right],
\label{f3cdmmodel}
\end{eqnarray}
with
\begin{eqnarray}
\alpha=\frac{\Omega_{F0}}{1-(1+\frac{2}{\beta})e^{-\frac{1}{\beta}}}.
\end{eqnarray}
The $\beta$ parameter is observationally constrained within the range 
$0.02<\beta<0.2$  \cite{Nesseris:2013jea,Nunes:2018evm,Anagnostopoulos:2019miu} 
and GR is recovered for 
$\beta\rightarrow0^+$.

The respective background and perturbation equations for the $f(T)$ models mentioned above are shown in \App{app:power_law_f_T} and in \App{app:exponential_f_T}.

\subsection{Inflation realization}

Regarding the choice of the inflationary potential we will work with  
inflationary setups with inflection points giving rise to an ultra slow-roll 
(USR) phase. In particular, during this USR phase, the non-constant mode of the 
curvature perturbations, which would otherwise decay exponentially in the 
slow-roll regime, in the USR phase will grow enhancing in this way the curvature 
power spectrum at specific scales which can potentially collapse forming 
PBHs. For concreteness, we will work within $\alpha$-attractor inflationary 
models~\cite{Dalianis:2018frf} naturally motivated by supergravity 
setups~\cite{Kallosh:2010ug}. In particular, we will work with the chaotic 
inflationary model which reads as 
\beq\label{eq:modulated_chaotic_potential}
V(\phi) = V_0\left\{\tanh\left(\frac{\phi} {\sqrt{6\alpha}}\right) + 
A_\mathrm{\phi}\sin\left[\tanh\left(\frac{\phi}{\sqrt{6\alpha}}
\right)/f_\phi\right] \right\}^2,
\eeq
as well as with the polynomial inflationary superpotential given by
\beq\label{eq:poly_superpotential}
\begin{split}
V(\phi)  = &  V_0\Bigl[c_0+c_1\tanh\left(\frac{\phi} {\sqrt{6\alpha}}\right) + 
c_2\tanh^2\left(\frac{\phi}{\sqrt{6\alpha}}\right) \\ & + 
c_3\tanh^3\left(\frac{\phi}{\sqrt{6\alpha}}\right)\Bigr]^2.
\end{split}
\eeq

Regarding the  values of $\alpha$, $V_0$, $A_\phi$, $f_\phi$, $c_0$, $c_1$, 
$c_2$, $c_3$ we used the fiducial values used in~\cite{Dalianis:2018frf} giving 
rise to an enhanced power spectrum at very small scales compared to the ones 
probed by CMB measurements. These fiducial values are given in the following 
Table 1 in units of $\Mp$.
\begin{table}[h!]
    \begin{center}
        \begin{tabular}{ |p{0.3cm}||p{1.2cm}|p{1.2cm}|p{1.3cm}|p{1cm}|p{0.5cm}|p{1cm}|p{1cm}|  }
            \hline
            \multicolumn{8}{|c|}{Table 1} \\
            \hline
            $\alpha$  & $A_\phi$ & $f_\phi$ & $V_0$ & $c_0$ & $c_1$ & $c_2$ & $c_3$ \\ 
            \hline
            $1$  & $0.130383$  & $0.129576$ & $2\times 10^{-10}$ & $ 0.16401$ & $0.3$ & $-1.426$ & $2.20313$ \\ 
            \hline
        \end{tabular} \label{Table1}
    \end{center}
\end{table}
{\vspace *{0.3 cm}}

At the end, as it was checked numerically,  our quantitative results discussed 
in subsection \ref{sec:GW_source_effect} turn to be independent 
of the choices of the aforementioned inflationary parameters.

\section{Scalar induced gravitational waves in $f(T,\phi)$ gravity  }
\label{SIGW}

In the previous section we have considered only the first-order scalar  and 
tensor perturbations. Here, we perturb the tensor part of the metric up to 
second order in order to extract the second order tensor perturbations induced 
by first order scalar perturbations working in terms of metric variables instead 
of the tetrad fields
\footnote{It is important to note that all the equations for 
the evolution of the scalar and tensor perturbations are independent of the 
choice of the formulation of the gravity theory, namely either in terms of the 
metric or in terms of the tetrad fields. Equivalently, we could have chosen to 
use appropriate tetrad fields that correspond to the line element (\ref{metric 
decomposition with tensor perturbations}) as for instance is done in 
\cite{Chen:2010va,Bahamonde:2021gfp}. }, which simplifies a lot the derivation 
of the tensor power spectrum and the GW signal.

\subsection{The scalar induced tensor 
perturbations}\label{sec:tensor_perturbations}

Working therefore within the Newtonian gauge frame with $\Phi=\Psi$ 
\footnote{ 
We can make this approximation since the anisotropic stress $\Pi$ is negligible for the time period we investigate; hence from the field equations \Eq{FieldEquations} [See also \cite{Cai:2015emx} for more details], $(1+F_{T})(\Psi-\Phi) = 8{\pi}G 
~\bar{p}\Pi~ \Rightarrow \Phi \approx \Psi$, where $F_T\equiv f(T)-T$.}, the perturbed 
Friedmann-Lema\^itre-Robertson-Walker metric 
~\cite{Ananda:2006af,Baumann:2007zm,Kohri:2018awv,Espinosa:2018eve,
CANTATA:2021ktz}, the perturbed metric can be written as 
\footnote{ We need to 
stress here 
that the gauge dependence of the tensor modes disappears in the case of scalar 
induced gravitational waves generated during a radiation-dominated era, as the 
one we focus on 
here, due to diffusion damping which exponentially suppresses the curvature 
perturbations in the late-time limit~\cite{DeLuca:2019ufz, Yuan:2019fwv, 
Inomata:2019yww,PhysRevLett.113.061301}.}
\begin{eqnarray}
\label{metric decomposition with tensor perturbations}
\begin{split}
\mathrm{d}s^2 = & a^2(\eta)\Biggl\lbrace-(1+2\Phi)\mathrm{d}\eta^2  \\ & 
+  
\left[(1-2\Phi)\delta_{ij} + 
\frac{h_{ij}}{2}\right]\mathrm{d}x^i\mathrm{d}x^j\Biggr\rbrace, 
\end{split}
\end{eqnarray}
where $\Phi$ is  the first order scalar perturbation, usually called as Bardeen 
potential, and $h_{ij}$ is the second-order tensor perturbation. Let us highlight 
here that we do not include in the analysis the contribution from the first 
order tensor perturbations since we focus on gravitational waves generated by 
scalar perturbations at second order. 

Working now  in the Fourier space, the equation of motion for the tensor 
perturbations $h_\boldmathsymbol{k}$ can be recast in the following 
form:~\cite{Ananda:2006af,Baumann:2007zm,Kohri:2018awv}
\beq
\label{Tensor Eq. of Motion}
h_\boldmathsymbol{k}^{\lambda,\prime\prime} + 
2\mathcal{H} (1-\gamma_T)  h_\boldmathsymbol{k}^{\lambda,\prime} + k^{2}  
h^\lambda_\boldmathsymbol{k} = 4 S^\lambda_\boldmathsymbol{k}\, ,
\eeq
where 
$\lambda = (+), (\times)$ and the source term $S^\lambda_\boldmathsymbol{k}$ reads as:
\beq
\label{Source}
S^\lambda_\boldmathsymbol{k}  =
\int\frac{\mathrm{d}^3 q}{(2\pi)^{3/2}} 
e^{\lambda}(\boldmathsymbol{k},\boldmathsymbol{q})F(\boldmathsymbol{q},
|\boldmathsymbol{k-q}|,\eta)\phi_\boldmathsymbol{q}\phi_\boldmathsymbol{k-q},
\eeq
with $e^{\lambda}(\boldmathsymbol{k},\boldmathsymbol{q}) \equiv e^s_{ij}(\boldmathsymbol{k})q_iq_j$ and the polarisation  tensors $e^{(+)}_{ij}$ and $e^{(-)}_{ij}$ being
defined as
\begin{eqnarray}
e^{(+)}_{ij}(\boldmathsymbol{k}) \equiv \frac{1}{\sqrt{2}} 
\left[e_i(\boldmathsymbol{k})e_j(\boldmathsymbol{k}) - 
\bar{e}_i(\boldmathsymbol{k})\bar{e}_j(\boldmathsymbol{k})\right], \\ 
e^{(\times)}_{ij}(\boldmathsymbol{k}) \equiv \frac{1}{\sqrt{2}} 
\left[e_i(\boldmathsymbol{k})\bar{e}_j(\boldmathsymbol{k}) + 
\bar{e}_i(\boldmathsymbol{k})e_j(\boldmathsymbol{k})\right],
\end{eqnarray}
where $e_i(\boldmathsymbol{k})$  and $\bar{e}_i(\boldmathsymbol{k})$ are two 
three-dimensional vectors which together with $\boldmathsymbol{k}/k$ form an 
orthonormal basis. In \Eq{Source}, the Fourier component of the Bardeen 
potential has been written as $\Phi_k(\eta) = T_\Phi(x)\phi_k$ with $x=k\eta$, 
where $\phi_k$ is the value of $\Phi$ at some reference initial time $x_0$, here 
considered as the horizon crossing time, and $T_\Phi(x)$ is a transfer function, 
defined as the ratio of the dominant mode of $\Phi$ between the times $x$ and 
$x_0$. Regarding the time evolution of $\Phi_k(\eta)$ this will be given from 
the time-time perturbed field equation within the torsional formulation of 
gravity, which in the absence of entropic perturbations reads like in 
GR~\cite{Papanikolaou:2022hkg} as
\beq
\label{Bardeen potential}
\Phi_k^{\prime\prime} + 
\frac{6(1+w)}{1+3w}\frac{1}{\eta}\Phi_k^{\prime} + w k^2\Phi_k =0\, .
\eeq
Finally, the function $F(\boldmathsymbol{q},|\boldmathsymbol{k-q}|,\eta)$ is 
defined in terms of the transfer function as
\begin{eqnarray}
\label{F}
\begin{split}
F(\boldmathsymbol{q}, &|\boldmathsymbol{k-q}|,\eta) \equiv 2T_\Phi(q\eta) 
T_\Phi\left(|\boldmathsymbol{k}-\boldmathsymbol{q}|\eta\right) 
\\ & + \frac{4}{3(1+w)}\left[\mathcal{H}^{-1}qT_\Phi^{\prime}(q\eta)+T_\Phi(q\eta)\right]\\  & \times \left[\mathcal{H}^{-1}\vert\boldmathsymbol{k}-\boldmathsymbol{q}\vert T_\Phi^{\prime}\left(|\boldmathsymbol{k}-\boldmathsymbol{q}|\eta\right)+T_\Phi\left(|\boldmathsymbol{k}-\boldmathsymbol{q}|\eta\right)\right].
\end{split}
\end{eqnarray}

Now, \Eq{Tensor Eq. of Motion} can be solved by virtue of the Green's function 
formalism with $h_\boldmathsymbol{k}^{s}$ being read as
\beq
\label{tensor mode function}
h^\lambda_\boldmathsymbol{k} (\eta)   =\frac{4}{a(\eta)} 
\int^{\eta}_{\eta_\mathrm{d}}\mathrm{d}\bar{\eta}\,  
G^\lambda_\boldmathsymbol{k}(\eta,\bar{\eta})a(\bar{\eta}
)S^\lambda_\boldmathsymbol{k}(\bar{\eta}),
\eeq
where the Green's function  $G^\lambda_{\bm{k}}(\eta,\bar{\eta})$ is the solution of the homogeneous equation 
\beq
\label{eq:Green function equation in f(T)}
\begin{split}
G_\boldmathsymbol{k}^{\lambda,\prime\prime}(\eta,\bar{\eta}) &  - 2\mathcal{H} \gamma_T 
G_\boldmathsymbol{k}^{\lambda,\prime}(\eta,\bar{\eta}) + \\ & \left( k^{2}  
-\frac{a^{\prime\prime}}{a}+ 2 \mathcal{H}^2 \gamma_T 
\right)G^\lambda_\boldmathsymbol{k}(\eta,\bar{\eta}) 
= 
\delta\left(\eta-\bar{\eta}\right),
\end{split}
\eeq
with the boundary conditions $\lim_{\eta\to 
\bar{\eta}}G^\lambda_\boldmathsymbol{k}(\eta,\bar{\eta}) = 0$ and $ \lim_{\eta\to 
\bar{\eta}}G^{\lambda,\prime}_\boldmathsymbol{k}(\eta,\bar{\eta})=1$. 

At the end, one can extract the tensor power spectrum $\mathcal{P}_h(k)$  
defined as the equal time correlation function of the tensor perturbations as 
follows:
\begin{eqnarray}
\label{tesnor power spectrum definition}
\langle h^\lambda_{\boldmathsymbol{k}_1}(\eta)h^{\rho,*}_{\boldmathsymbol{k}_2}(\eta)\rangle \equiv \delta^{(3)}(\boldmathsymbol{k}_1 - \boldmathsymbol{k}_2) \delta^{\lambda\rho} \frac{2\pi^2}{k^3_1}\mathcal{P}^{(\lambda)}_{h}(\eta,k_1),
\end{eqnarray}
where $\lambda=(\times)$ or $(+)$. After a long but straightforward  calculation 
and accounting for the fact that on the superhorizon regime 
$\Phi=2\mathcal{R}/3$~\cite{Mukhanov:1990me}, where $\mathcal{R}$ is the 
comoving curvature perturbation, $\mathcal{P}_h(k)$ can be recast 
as~\cite{Kohri:2018awv,Espinosa:2018eve}
\begin{eqnarray}
\label{Tensor Power Spectrum}
\begin{split}
\mathcal{P}^{(\lambda)}_h(\eta,k) = 4\int_{0}^{\infty} & \mathrm{d}v\int_{|1-v|}^{1+v}\mathrm{d}u \left[ \frac{4v^2 - (1+v^2-u^2)^2}{4uv}\right]^{2}\\ & \times I^2(u,v,x)\mathcal{P}_\mathcal{R}(kv)\mathcal{P}_\mathcal{R}(ku)\,,
\end{split}
\end{eqnarray}
with 
\begin{eqnarray}
\label{I function}
I(u,v,x) = \int_{x_0}^{x} \mathrm{d}\bar{x}\, \frac{a(\bar{x})}{a(x)}\, k\, G_{k}(x,\bar{x}) F_k(u,v,\bar{x}).
\end{eqnarray}

\subsection{The gravitational wave spectral abundance}

 Finally, defining the effective energy density of the gravitational waves in 
the subhorizon region where one can use the flat spacetime approximation and 
where \Eq{Tensor Eq. of Motion} reduces to a free-wave equation, one can 
straightforwardly show [See~\cite{Maggiore:1999vm,Isaacson:1968zza} for more 
details] that the GW spectral abundance $\OmegaGW$ defined as the GW energy 
density contribution per logarithmic comoving scale, will read as
\beq\label{Omega_GW}
\Omega_\mathrm{GW}(\eta,k)\equiv \frac{1}{\bar{\rho}_\mathrm{tot}} 
\frac{\mathrm{d}\rho_\mathrm{GW}(\eta,k)}{\mathrm{d}\ln k} = 
\frac{1}{24}\left(\frac{k}{\calH(\eta)}\right)^{2}\overline{\mathcal{P}^{
(\lambda)}_h(\eta,k)},
\eeq
with the bar standing for an averaging over the sub-horizon oscillations of the tensor field, which is done in order to only extract the envelope of the GW spectrum at those scales.
 
One then can account for the Universe expansion and derive the GW energy density 
 contribution today. In order to achieve that, one writes   
\beq
\begin{split}
\OmegaGW(\eta_0,k) & = \frac{\rhoGW(\eta_0,k)}{\rho_\mathrm{c}(\eta_0)} = 
\frac{\rhoGW(\eta_\mathrm{*},k)}{\rho_\mathrm{c}(\eta_\mathrm{*})}\left(\frac{
a_\mathrm{*}}{a_\mathrm{0}}\right)^4 
\frac{\rho_\mathrm{c}(\eta_\mathrm{*})}{\rho_\mathrm{c}(\eta_0)}\\ & = 
\OmegaGW(\eta_\mathrm{*},k)\Omega^{(0)}_\mathrm{r}\frac{\rho_\mathrm{r,*}
a^4_\mathrm{*}}{\rho_\mathrm{r,0}a^4_0},
\end{split}
\eeq
where  the  index 
$0$ denotes our present time and $\eta_\mathrm{*}$ is a reference time usually  
taken as the horizon crossing time when one considers that an enhanced energy 
perturbation with a characteristic scale $k$ collapses to form a PBH. For the 
above expression, we accounted for the fact that $\Omega_\mathrm{GW}\sim 
a^{-4}$. Then, using the fact that the energy 
density of radiation reads as $\rho_r = 
\frac{\pi^2}{15}g_{*\mathrm{\rho}}T_\mathrm{r}^4$ and that the temperature of 
the radiation bath, $T_\mathrm{r}$, scales as $T_\mathrm{r}\propto 
g^{-1/3}_{*\mathrm{S}}a^{-1}$, one acquires that 
\beq\label{Omega_GW_RD_0}
\Omega_\mathrm{GW}(\eta_0,k) = 
\Omega^{(0)}_r\frac{g_{*\mathrm{\rho},\mathrm{*}}}{g_{*\mathrm{\rho},0}}
\left(\frac{g_{*\mathrm{S},\mathrm{0}}}{g_{*\mathrm{S},\mathrm{*}}}\right)^{4/3}
\OmegaGW(\eta_\mathrm{*},k),
\eeq
where $g_{*\mathrm{\rho}}$ and $g_{*\mathrm{S}}$ stand for the energy and 
entropy relativistic degrees of freedom.

\subsection{Teleparallel gravity modifications of the gravitational wave signal}

Having extracted before the SIGW signal within modified teleparallel theories of gravity  
we investigate here the relevant modifications of $f(T)$ theories at the level 
of the source and the propagation of the GWs which can potentially render the GW 
distinctive with respect to the one within classical gravity.  

\subsubsection{The effect at the level of the gravitational wave 
source}\label{sec:GW_source_effect}

Regarding the effect of the underlying modified teleparallel gravity theory at the level 
of GW source, it will be encapsulated in the curvature power spectrum, which 
 actually constitutes the source of the SIGWs as it can be inferred from \Eq{Tensor Power Spectrum} and
\Eq{Omega_GW}.

Working within the framework of the mono-parametric $f(T)$ models  introduced 
previously we solve numerically the Mukhanov-Sasaki equation and extract the 
curvature spectrum at the end of inflation on super-horizon scales, which is 
actually what will induce the tensor power spectrum seen in \Eq{Tensor Power 
Spectrum}. For our numerical applications we choose to work within the framework 
of $\alpha$ attractor inflationary potentials introduced in \Sec{sec:f(T) 
models} and which present an inflection point behavior necessary for the 
enhancement of the curvature perturbations on small scales.

\begin{figure}[ht]
\begin{center}
\includegraphics[width=0.496\textwidth, clip=true]
{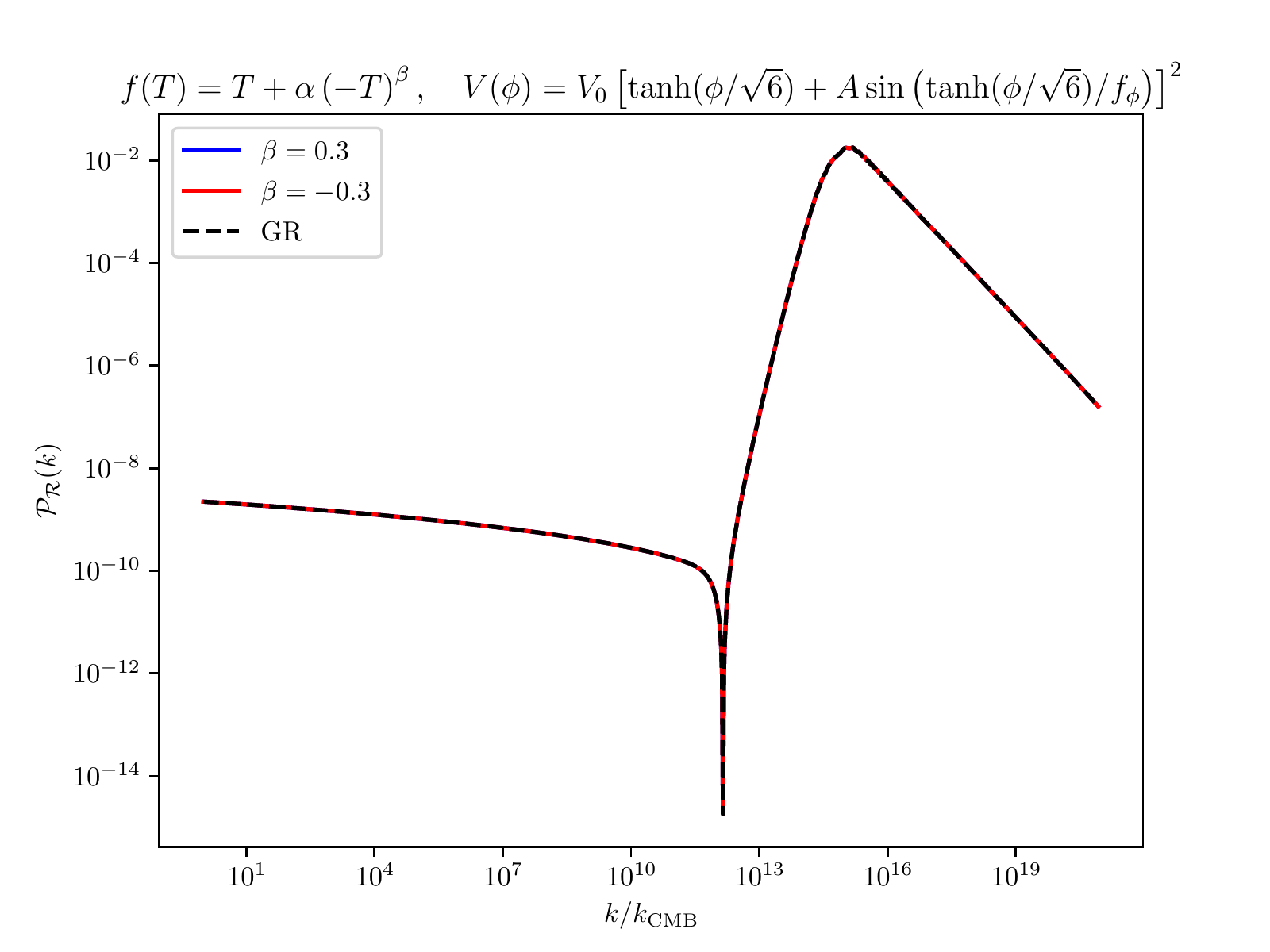}
\includegraphics[width=0.496\textwidth, clip=true]{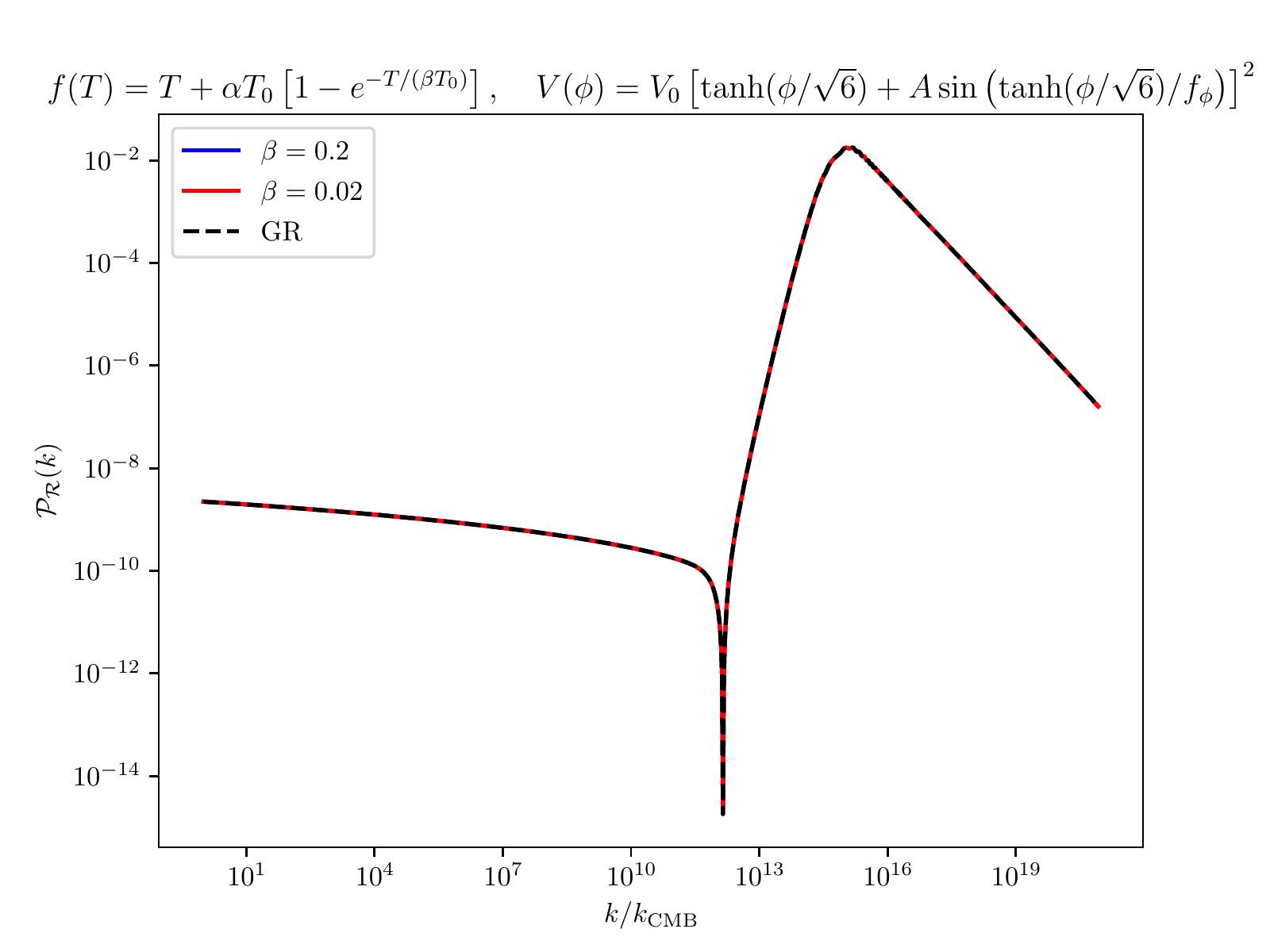}
\end{center}
\caption{{\it{In top panel we show the curvature power spectrum 
$\mathcal{P}_\mathcal{R}(k)$ for the power-law $f(T)$ model for various values 
of the $\beta$ parameter while in the bottom panel we show 
$\mathcal{P}_\mathcal{R}(k)$ for the exponential $f(T)$ model. The black dashed 
line stands for $\mathcal{P}_\mathcal{R}(k)$ within GR. For all curves we work 
with the modulated chaotic inflationary potential 
(\ref{eq:modulated_chaotic_potential}).}}}
\label{fig:modulated_chaotic_potential}
\end{figure}

\begin{figure}[ht]
\begin{center}
\includegraphics[width=0.496\textwidth, clip=true]
{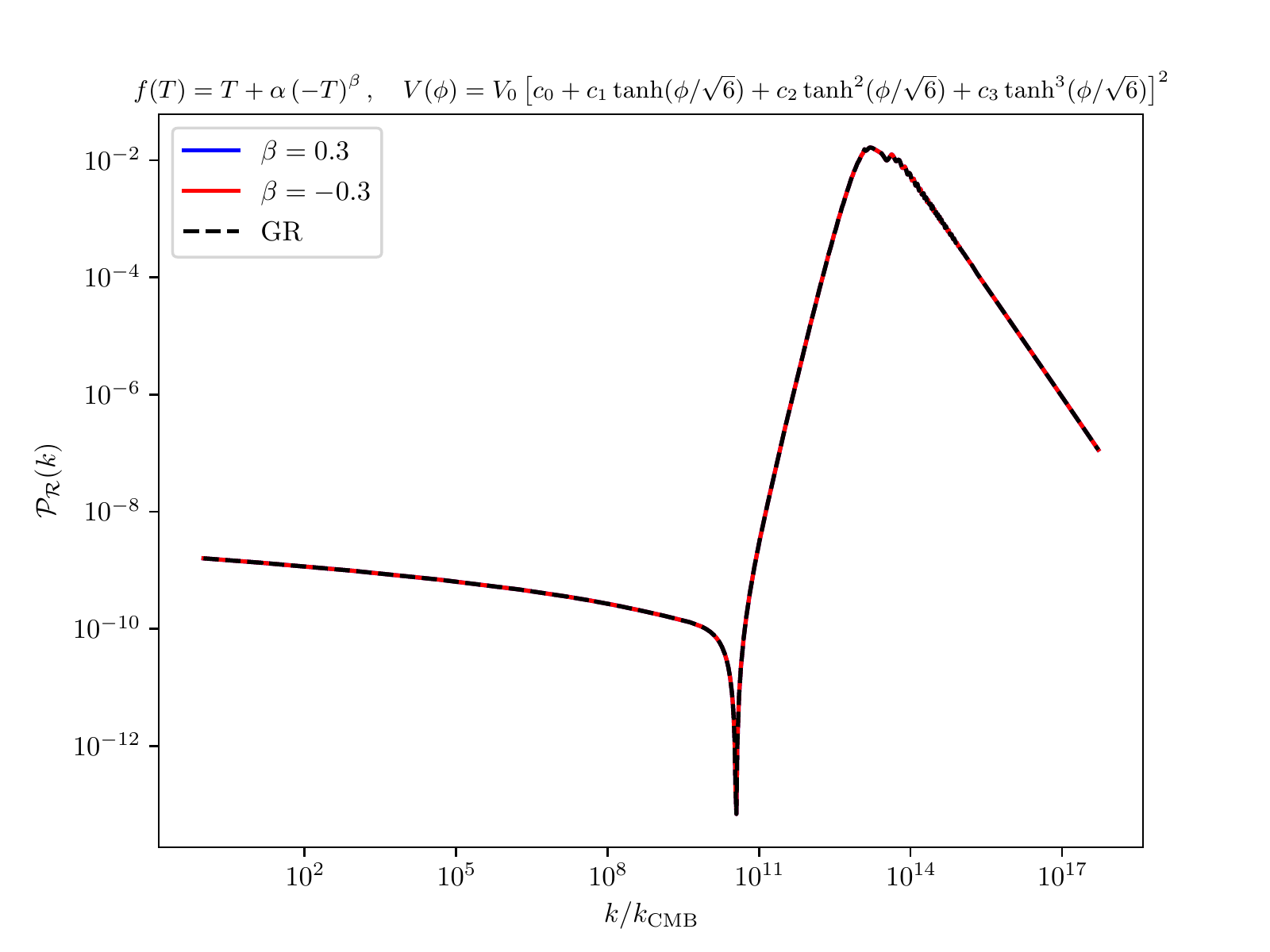}
\includegraphics[width=0.496\textwidth, clip=true]{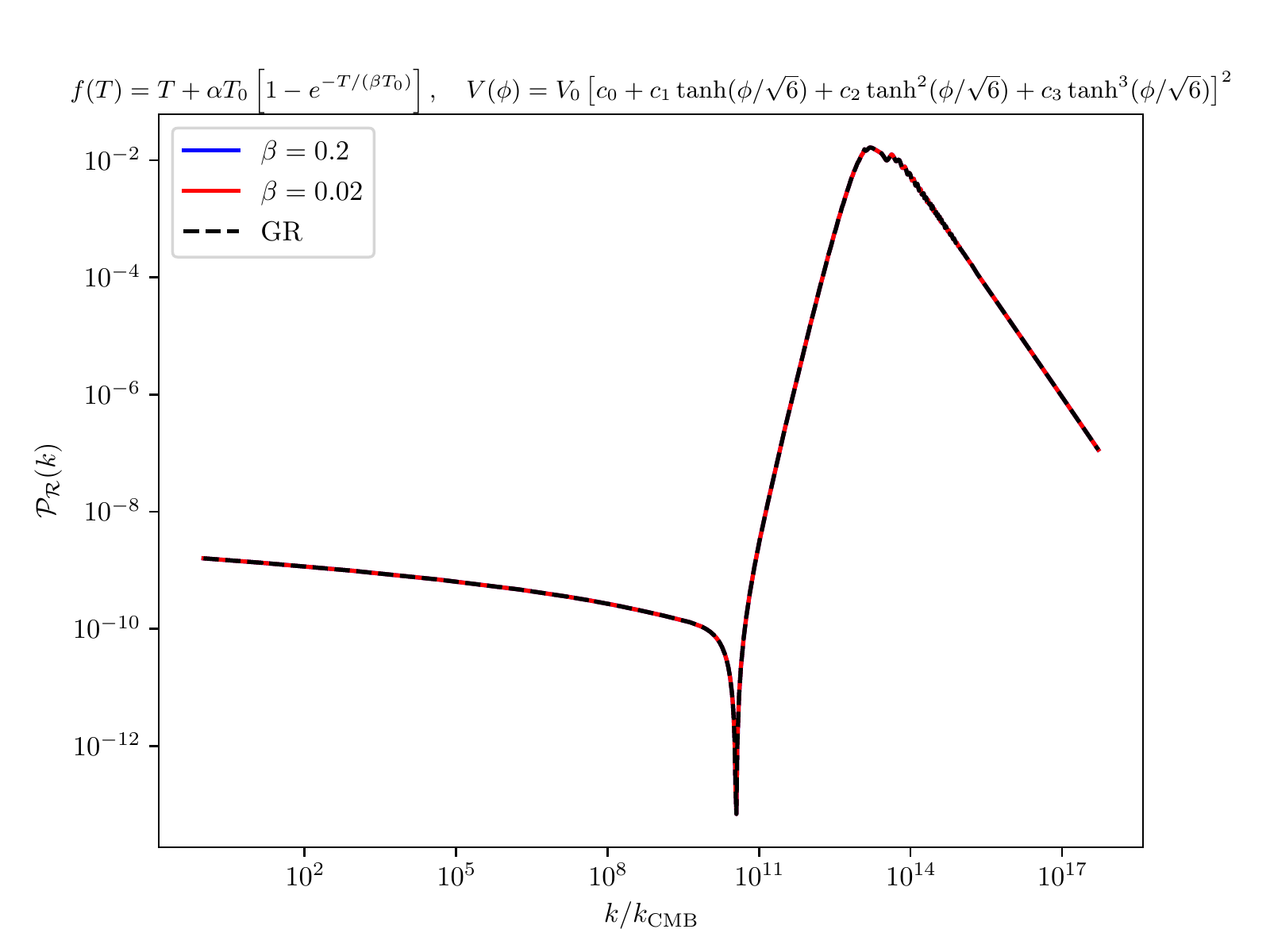}
\end{center}
\caption{{\it{In the top panel we show the curvature power spectrum 
$\mathcal{P}_\mathcal{R}(k)$ for  the power-law $f(T)$ model for various values 
of the $\beta$ parameter while in the bottom panel we show 
$\mathcal{P}_\mathcal{R}(k)$ for the exponential $f(T)$ model. The black dashed 
line stands for $\mathcal{P}_\mathcal{R}(k)$ within GR. For all curves we work 
with the polynomial superpotential (\ref{eq:poly_superpotential}).}}}
\label{fig:poly_superpotential}
\end{figure}

In  \Fig{fig:modulated_chaotic_potential} we show the curvature 
power spectrum for the case of the modulated chaotic inflationary potential 
(\ref{eq:modulated_chaotic_potential}) and the two power-law and the exponential 
$f(T)$ models by varying the modified gravity parameter $\beta$ within its 
observationally allowed range. In \Fig{fig:poly_superpotential} we show the respective $\mathcal{P}_\mathcal{R}(k)$ for the case of the polynomial 
superpotential (\ref{eq:poly_superpotential}). As it can be observed for both 
figures, the curvature power spectrum derived within modified teleparallel gravity 
theories is practically indistinguishable from that of classical gravity with 
the relative difference of $\mathcal{P}_\mathcal{R}(k)$ with respect to the one 
of GR being of the order  $10^{-18}$, namely
\beq
\left\vert\frac{\mathcal{P}^{f(T)}_\mathcal{R}(k) 
-\mathcal{P}^{\mathrm{GR}}_\mathcal{R}(k) 
}{\mathcal{P}^{\mathrm{GR}}_\mathcal{R}(k)}\right\vert\sim 10^{-18}.
\eeq

At this point, we need to highlight that we derived the curvature power spectrum 
within mono-parametric $f(T)$ gravity theories without non-minimal matter-gravity couplings. 
One in general would expect a different behaviour in the case where there is a non-minimal 
coupling between the gravity and the matter sector, namely when $f_{,T\phi}\neq 
0$. This intuitive physical condition can be analytically derived from extracting 
$\mathcal{P}_\mathcal{R}(k)$ within the slow-roll regime and checking which are the necessary conditions for the curvature power spectrum 
within teleparallel gravity to be distinctive from that within GR.  

For this reason, let us derive here the $\mathcal{P}_\mathcal{R}(k)$ at linear order 
in the slow roll regime, namely when $\epsilon,\eta\ll 1$, within the framework 
of $f(T)$ gravity theories
\footnote{Strictly speaking, \Eq{eq:R_k_SR} and 
\Eq{eq:P_RR_SR} are partially valid within the ultra slow-roll inflationary 
regime, since they are not valid at all scales. In particular, during USR inflation 
curvature perturbations do not freeze out at horizon exit time and 
therefore \Eq{eq:R_k_SR} and \Eq{eq:P_RR_SR} cannot be evaluated at horizon 
crossing time but rather only after USR inflation ends. In our case, we derive 
$\mathcal{P}_\mathcal{R}(k)$  at the end of inflation, so to that 
end, \Eq{eq:R_k_SR} and \Eq{eq:P_RR_SR} extracted within the SR regime can be 
used as a first approximation for the curvature power spectrum. 
See~\cite{Byrnes:2021jka} for a more detailed discussion.}. 
After appropriate 
approximations (see \cite{Gonzalez-Espinoza:2020azh} for more details), we can 
write
\beq\label{eq:R_k_SR}
\begin{split}
\left|\mathcal{R}_{k}\right| \simeq & \frac{H}{2 \sqrt{ k^3 Q_{s}}} 
\left(\frac{k}{a H}\right)^{\frac{3}{2}-\tilde{\nu}} \\ & \simeq \frac{H_{k}}{ 2 
\sqrt{ k^3 Q_{sk}}}\left[1+\eta_{\mathcal{R}}\ln\left(\frac{k}{a 
H}\right)\right],
\end{split}
\eeq
where $H_{k}$ and $Q_{sk}$ are the values of $H$ and $Q_{s}$ evaluated at 
horizon  crossing time, namely when $k=a H$.
Finally, the curvature power spectrum can be recast as
\begin{eqnarray}\label{eq:P_RR_SR}
\mathcal{P}_\mathcal{R}(k) \equiv \frac{k^3}{2 \pi^2} 
\left|\mathcal{R}_{k}(\tau)\right|^2 \simeq \frac{H_{k}^2}{8 \pi^2 
Q_{sk}}\left[1+2\eta_{\mathcal{R}}\ln\left(\frac{k}{a H}\right)\right].
\end{eqnarray} 

Given that $\eta_{\mathcal{R}}\sim \mathcal{O}(\epsilon)$,  one finds that the 
local Lorentz violation gives rise to a slight logarithmic time-dependence of 
the curvature perturbation and its power spectrum on superhorizon scales. 
Finally, the scale-dependence of the curvature power spectrum is quantified 
in the scalar spectral index $n_\mathrm{s}$ defined by 
\begin{equation}
{n_\mathrm{s}-1\equiv \left.\frac{d \ln{\mathcal{P}_\mathcal{R}(k)}}{d\ln{k}}\right|_{k=a H}=-2\epsilon-\eta+2 \eta_{\mathcal{R}},
\label{ns_fTphi}}
\end{equation} 
from which we see the deviation from GR due to the  presence of the term $2 
\eta_{\mathcal{R}}$, which carries the effects of the local Lorentz 
violation. 

Finally, one can show that $\eta_\mathcal{R}$ can be recast in the following form:
\beq
\eta_\mathcal{R} = (\delta_{fX} - 2\mu \epsilon)\left[1 -  
\frac{1+2\mu}{(1+2\mu)\epsilon-\delta_{fX}}\frac{\delta_{fX}-2\mu\epsilon}{2\mu}
\right].
\eeq
Interestingly, for $\delta_{fX}=0$, i.e. in the absence of matter-gravity  
coupling, $\eta_\mathcal{R}=-4\mu\epsilon\ll 1$, since
$\epsilon<1$ and $\mu = Tf_{,TT}/f_{,T}\ll 1$ for 
viable $f(T)$ models \cite{Nesseris:2013jea} and $Q_{sk} = \epsilon_k$. Thus, in the absence of matter-gravity coupling one 
obtains that $\mathcal{P}^{f(T)}_\mathcal{R}(k)\simeq 
\mathcal{P}^{\mathrm{GR}}_\mathcal{R}(k)$ and consequently can claim 
that there will be essentially no distinctive deviation between $f(T)$ and GR at the 
level of the curvature power spectrum.

One therefore should introduce a matter-gravity coupling at the level of the  
Lagrangian in order to detect a potential deviation from GR at the level of 
$\mathcal{P}_\mathcal{R}(k)$ constituting the source of the SIGWs.

\subsubsection{The effect of the gravitational-wave propagation}

We now study the effect of the underlying telleparallel gravity theory at the 
level  of the GW propagation. To do so, one should essentially investigate the 
behavior 
of the Green function, $G_k(\eta,\bar{\eta})$, which can be viewed as the propagator of the 
tensor perturbations as it can be seen from \Eq{tensor mode function}.

In particular, one must identify the dominant terms in the evolution equation 
for the Green  function \Eq{eq:Green function equation in f(T)}, which we write 
as follows
\beq\label{eq:Green_function_propagation_effect}
\begin{split}
G_\boldmathsymbol{k}^{\lambda,\prime\prime}(\eta,\bar{\eta})  - &  2\mathcal{H} \gamma_T 
G_\boldmathsymbol{k}^{\lambda,\prime}(\eta,\bar{\eta}) + \\ & \left( k^{2}  
-\frac{a^{\prime\prime}}{a}+ 2 \mathcal{H}^2 \gamma_T 
\right)G^\lambda_\boldmathsymbol{k}(\eta,\bar{\eta}) 
= 
\delta\left(\eta-\bar{\eta}\right),
\end{split}
\eeq
and take the 
ratios between 
the GR terms and the new $f(T)$
terms multiplied by the 
$\gamma_T$ function.

At the end, following the same reasoning as in~\cite{Papanikolaou:2022hkg} and 
accounting  for the fact that 
the $\gamma_T$ 
function in the case of no non-minimal matter-gravity coupling is a negative decreasing 
function of time,  we derive the maximum 
deviation from GR by extracting the ratios between the GR and $f(T)$ terms at a 
time during radiation domination when the $\gamma_T$ function acquires its 
maximum value. Being quite 
conservative, we choose this time as the standard matter-radiation equality time 
at redshift $z_\mathrm{eq}=3387$. Finally, we find that independently of the 
value of the $f(T)$ gravity parameter $\beta$ one obtains that 

\beq
\begin{split}
& \left\vert\frac{G_\boldmathsymbol{k}^{\prime\prime}(\eta,\bar{\eta})}{2\mathcal{H} 
\gamma_T G_\boldmathsymbol{k}^{\prime}(\eta,\bar{\eta})}\right\vert \simeq 
\left.\frac{1}{2\mathcal{H} \gamma_T}\right\vert_{\eta=\eta_\mathrm{eq}} 
\gg 1\quad \mathrm{and} \\ & \left. \frac{k^2}{2\mathcal{H}^2 \gamma_T 
}\right\vert_{k=k_\mathrm{evap},\eta=\eta_\mathrm{eq}}\gg 1,
\end{split}
\eeq
where $k_\mathrm{evap}$ is the comoving scale exiting the Hubble radius at PBH 
evaporation  time, thus being the largest scale considered here. 

In summary, we can safely argue that the modifications of any modified 
teleparallel  gravity theory with no non-minimal gravity-matter coupling at the level of 
the GW propagation  equation \eqref{eq:Green_function_propagation_effect} are 
negligible. As a consequence, one concludes that
\beq
G_\boldmathsymbol{k}^{f(T)}
(\eta,\bar{\eta})\simeq
G_\boldmathsymbol{k}^{\mathrm{GR}}(\eta,{\bar{\eta}}),\quad\mathrm{with}\quad f_{,T\phi}=0.
\eeq
One then needs to  introduce a coupling between gravity and matter in order to 
see a distinctive deviation from GR.

\section{Conclusions}
\label{sec:Conclusions}

Primordial black holes are of great significance, since they can naturally 
address many issues of modern cosmology, among them the dark matter problem and 
the generation of large-scale structures. Interestingly, they are associated 
with numerous GW signals from GWs from PBH mergers up to primordial GWs of 
cosmological origin related to their formation.

In particular,  the enhanced cosmological perturbations which collapse to form 
PBHs can induce a stochastic gravitational-wave background due to second-order 
gravitational interactions. This GW portal was mainly studied within classical 
gravity while in some early works in this research area it was shown that it can 
as well serve as a novel probe to test and constrain alternative gravity 
theories.

In this work,  we studied the aforementioned GW signal within the context of 
modified teleparallel gravity theories where the gravitational Lagrangian is a 
function of the torsion scalar $T$. Interestingly enough, we showed that in the 
absence of explicit non-minimal couplings between   gravity and   matter 
sectors, the effect of the underlying modified theory of gravity at the level of 
the source and the propagation of the GWs is practically negligible, leading to 
an indistinguishable SIGW signal compared to that within GR. Additionally, 
we would like to mention here that a similar indistinguishable GW signal 
compared to GR was found as well regarding the GW portal associated to
PBH Poisson fluctuations within teleparallel theories of 
gravity~\cite{Papanikolaou:2022hkg}.

Finally, it is important to highlight that one needs to introduce a non-minimal matter-gravity couplings in order to 
observe a distinctive SIGW signal compared to GR. Furthermore, it would  be 
illuminating to extract the induced GW signal within other modified 
gravity theories, namely within $f(R)$ and $f(Q)$ gravity theories, in order to 
potentially test and constrain the underlying theory of gravity. These studies 
will be performed in upcoming projects.

\begin{acknowledgments}
C.T. and T.P. acknowledge financial support from the A.G. Leventis Foundation  
and the Foundation for Education
and European Culture in Greece respectively.  The authors would also like to 
acknowledge the 
contribution of the COST Actions  
CA18108 ``Quantum Gravity Phenomenology in the multi-messenger approach''  and 
CA21136 ``Addressing observational tensions in cosmology with systematics and 
fundamental physics (CosmoVerse)''.

\end{acknowledgments}

\appendix

\section{The power-law $f(T)$ model}\label{app:power_law_f_T}
 \subsection{Background equations}

For the power-law $f(T)$ model (\ref{powermod}), after assuming homogeneity and 
isotropy of the scalar field (hence $X = \dot{\phi}^2/2$) and working with the 
e-fold number defined as the logarithm of the scale factor, i.e. $N\equiv \ln 
a$, Eqs. (\ref{Fr1}), (\ref{Fr2}) and (\ref{phi}) become
\beq
\label{Fr11}
\frac{T \Mp^2}{2} +  \frac{T^\beta \Mp^2 \alpha (2\beta -1)}{2} - \frac{H^2\phi^{\prime 2}}{2} - V =0, 
\eeq
\beq
\label{Fr22}
\begin{split}
\frac{T \Mp^2}{2}+ & \frac{T^\beta \Mp^2 \alpha (2\beta -1)}{2} + \frac{H^2\phi^{\prime 2}}{2} -V -2\epsilon H^2 \Mp^2  \\ & - 2\epsilon H^2 \Mp^2\alpha \beta \left[ T^{\beta -1} +(\beta -1) H T^{\beta -2}T^\prime\right] =0 
\end{split}
\eeq
\beq
\label{KG1}
\phi^{\prime\prime} + (3 -\epsilon)  \phi^\prime + V_{,\phi}/H^2=0 
\eeq
where $\prime$ denotes derivative with respect to the e-fold  number, $\epsilon 
= -H^\prime/H$, $T=6H^2$ and $T^\prime = 12H H^\prime$. 

\subsection{Mukhanov - Sasaki equation}
We need to solve these equations together with the Mukhanov - Sasaki equation  
(\ref{MK_conformal}). Working again with the e-fold number as time variable one 
obtains from \Eq{MK_conformal} that
\beq
\mathcal{R}^{\prime\prime} + \left( 1-  \epsilon + 2\frac{z'}{z} \right)\mathcal{R}^\prime + \left( \frac{k^2}{H^2 
a^2} + \frac{m^2}{H^2}\right)\mathcal{R} = 0, \label{MK2_app_power_law} 
\eeq
with 
\begin{eqnarray}
    \frac{z'}{z}= 1+ \phi ''/\phi', \, \, \, \, \epsilon = \frac{{\phi'}^2/(2M_{pl}^2)}{1+\alpha \beta (2\beta -1) T^{\beta -1}}.
\end{eqnarray}
The expression for $m^2$ will read as 
\beq
m^2=3 H^2 \eta_{\mathcal{R}} = -72H^4 \frac{\alpha \beta (\beta - 1) 
(6H^2)^{\beta-2} \epsilon}{1+ \alpha \beta (6H^2)^{\beta -1}} .
\eeq
 \vspace{1cm}
\section{The exponential $f(T)$ model}\label{app:exponential_f_T}

\subsection{Background equations}
For exponential model (\ref{f3cdmmodel}), following the same procedure as 
before (hence $X = \dot{\phi}^2/2$) and working with the e-fold number as the 
time variable equations (\ref{Fr1}), (\ref{Fr2}) and (\ref{phi}) become
\beq
\label{FR1F2}
\begin{split}
3H^2 \Mp^2 - 3H_0^2\Mp^2 \alpha  \left(1- e^{-\frac{H^2}{\beta H_0^2 }}\right) &  + \frac{ 6 H^2 \Mp^2 \alpha}{\beta}e^{\frac{-H^2}{\beta H_0^2}} \\ & - \frac{\phi'^2 H^2}{2} -V =0
\end{split}
\eeq
\beq
\label{FR2F2}
\begin{split}
3H^2 \Mp^2 & -3H_0^2\Mp^2 \alpha  \left(1- e^{-\frac{H^2}{\beta H_0^2 }}\right) +\frac{ 6 H^2 \Mp^2 \alpha}{\beta}e^{\frac{-H^2}{\beta H_0^2}} \\ & + \frac{\phi'^2 H^2}{2} -V - 2\epsilon H^2 \Mp^2\left( 1+ \frac{\alpha}{\beta}e^{\frac{-H^2} {\beta H_0^2}} \right)  \\ & + 4H^2 \Mp^2 e^{\frac{-H^2}{\beta H_0^2}} \frac{\alpha  \epsilon H^2}{\beta^2 H_0^2} =0
\end{split}
 \eeq
 \beq
\label{KGF2}
 \phi'' + (3 - \epsilon) \phi ' +  \frac{V_{,\phi}}{H^2} = 0. 
 \eeq 
 
\subsection{Mukhanov - Sasaki equation}

 Once again, regarding the MS equation one should solve the following  equation 
for the comoving curvature perturbation $\mathcal{R}$:
 \beq
\mathcal{R}'' + \left( 1- \epsilon + 2\frac{z'}{z} \right)\mathcal{R}' + \left( \frac{k^2}{H^2 a^2} + \frac{m^2}{H^2}\right)\mathcal{R} = 0, \label{MK2_app_exponential} 
\eeq
with 
\beq
\begin{split}
& \frac{z'}{z}= 1+ \phi ''/\phi',\quad \epsilon = 
\frac{{\phi'}^2/(2M_{pl}^2)}{1+\alpha \beta (2\beta -1) T^{\beta -1}}, \quad \\ & 
m^2= -12 H^2 \frac{ \epsilon \alpha H^2  e^{\frac{-H^2}{\beta H_0^2} }}{ \beta^2 
H_0^2 \left(1+\frac{\alpha} {\beta} e^{\frac{-H^2}{\beta H_0^2}} \right)}.
\end{split}
\eeq 

\bibliographystyle{JHEP} 
\bibliography{PBH}
\end{document}

%% file: newcommands.tex
\newcommand{\OmegaGW}{\Omega_{\mathrm{GW}}}
\newcommand{\rhoGW}{\rho_{\mathrm{GW}}}

\let\oldsqrt\sqrt
\def\sqrt{\mathpalette\DHLhksqrt}
\def\DHLhksqrt#1#2{%
\setbox0=\hbox{$#1\oldsqrt{#2\,}$}\dimen0=\ht0
\advance\dimen0-0.2\ht0
\setbox2=\hbox{\vrule height\ht0 depth -\dimen0}%
{\box0\lower0.4pt\box2}}

\newcommand{\sss}[1]{{\scriptscriptstyle{#1}}}
\newcommand{\boldmathsymbol}[1]{{\ensuremath{\boldsymbol{#1}}}}

\newcommand{\uPl}{\mathrm{Pl}}

\newcommand{\usssPl}{\sss{\uPl}}

\newcommand{\calH}{\mathcal{H}}

\newcommand{\Mp}{M_\usssPl}

\newcommand{\beq}{\begin{equation}}
\newcommand{\eeq}{\end{equation}}
\newcommand{\bea}{\begin{equation}\begin{aligned}}
\newcommand{\eea}{\end{aligned}\end{equation}}

\newlength{\wsingfig}
\newlength{\wdblefig}
\newlength{\wquadfig}
\newlength{\wtriplefig}
\newcommand{\Eq}[1]{Eq.~(\ref{#1})}

\newcommand{\Fig}[1]{Fig.~{\ref{#1}}}

\newcommand{\Sec}[1]{Sec.~\ref{#1}}

\newcommand{\App}[1]{Appendix~\ref{#1}}